%% file: stability251118.tex
\documentclass[11pt]{article}
\pdfoutput=1  
\usepackage{graphicx,color}
\usepackage{appendix}
\usepackage{latexsym,amsmath,amssymb,graphicx,booktabs}
\usepackage{epsfig,latexsym,cite}
\usepackage{hyperref}
\numberwithin{equation}{section}

\definecolor{MyBlue}{rgb}{0.15,0.15,0.70}

\hypersetup{
colorlinks=true,
citecolor=MyBlue,
linkcolor=MyBlue,
urlcolor=MyBlue
}

\input epsf
\usepackage{amssymb}
\usepackage{amsmath}
\usepackage{amsfonts}
\usepackage{upgreek}
\usepackage{latexsym}

\include{mydefs}

\begin{document}

\begin{titlepage}

\vspace*{2cm}

\centerline{\Large \bf Stability issues of nonlocal gravity during primordial inflation}

\vspace{5mm}


\vskip 0.4cm
\vskip 0.7cm
\centerline{\large Enis Belgacem$^a$, Giulia Cusin$^a$, Stefano Foffa$^a$, Michele Maggiore$^a$ and Michele Mancarella$^{b,c}$ }
\vskip 0.3cm
\centerline{\em $^a$D\'epartement de Physique Th\'eorique and Center for Astroparticle Physics,}  
\centerline{\em Universit\'e de Gen\`eve, 24 quai Ansermet, CH--1211 Gen\`eve 4, Switzerland}

\vspace{3mm}
\centerline{\em $^b$Institut de physique th\' eorique, Universit\'e  Paris Saclay}
\centerline{\em CEA, CNRS, 91191 Gif-sur-Yvette, France}
\vspace{3mm}
\centerline{\em $^c$Universit\'e Paris Sud, 15 rue George Cl\'emenceau, 91405,  Orsay, France}

\vskip 1.9cm

\begin{abstract}
We study the cosmological evolution of some nonlocal gravity models, when the initial conditions are set during a phase of primordial inflation. We examine in particular three models, the so-called RT, RR and $\Delta_4$ models, previously introduced by our group. We find that the RR and $\Delta_4$ models have a stable evolution also during inflation. The RT model has an apparent instability, but we show that, because of the smallness of the scale associated to  the nonlocal term compared to the inflationary scale, this instability is innocuous and also the RT model has  a viable evolution even when its initial conditions are set during a phase of primordial inflation.

\end{abstract}

\end{titlepage}

\newpage

\section{Introduction}

Understanding the origin of dark energy (DE)
is among the most important problems in cosmology. The simplest solution, a cosmological constant, fits the data very well, so  $\Lambda$CDM is at present the standard cosmological paradigm. However, the accuracy of present and future observations allows us  to test $\Lambda$CDM against competing theories, so it is clearly interesting to develop alternatives to $\Lambda$CDM. At the conceptual level, an especially intriguing possibility is to explain DE by modifying General Relativity (GR) at cosmological distances, see e.g.  \cite{DeFelice:2010aj,deRham:2014zqa,Schmidt-May:2015vnx} for  reviews of different approaches. 

Nonlocality opens up new interesting possibilities for building large-distance (``infrared")  modifications of GR.
While at the fundamental level quantum field theory is local, in the presence of massless or light particles the quantum effective action that includes quantum  corrections unavoidably develops nonlocal terms, both at the perturbative and at the non-perturbative level. In this spirit,  phenomenological nonlocal modifications of GR were  proposed in \cite{Wetterich:1997bz} and in
\cite{Deser:2007jk,Deser:2013uya} (see   \cite{Woodard:2014iga} for  review). 
In the last few years our group has introduced a different class of nonlocal models, characterized by the fact that the nonlocal terms are associated with a mass scale. These models, which evolved from previous work related to the degravitation idea~\cite{ArkaniHamed:2002fu,Dvali:2006su,Dvali:2007kt} as well as from attempts at writing massive gravity in nonlocal form~\cite{Porrati:2002cp,Jaccard:2013gla},  turn out to work extremely well in the comparison with cosmological observations. A first  model of this class, proposed in  \cite{Maggiore:2013mea},
is based on a nonlocal equation of motion,
\be\label{RT}
\Gmn -\frac{m^2}{3}\(\gmn\iBox R\)^{\rm T}=8\pi G\,\Tmn\, ,
\ee
where 
the superscript T denotes the operation of taking the transverse part of a tensor (which is itself a non-local operation), and ensures  energy-momentum conservation. 
We will refer to this model as the ``RT" model, where R stands for the Ricci scalar and T for the extraction of the transverse part.
A second model 
was introduced in \cite{Maggiore:2014sia}, and is defined by the quantum effective action
\be\label{RR}
\Gamma_{\rm RR}=\frac{\mplr^2}{2}\int d^{4}x \sqrt{-g}\, 
\[R-\frac{m^2}{6} R\frac{1}{\Box^2} R\]\, ,
\ee
where  $\mplr$ is the reduced Planck mass, $\mplr^2=1/(8\pi G)$, and again  $m$ is the new mass parameter of the model.
We will refer to it as the RR model. Finally, a third interesting model
is the ``$\Delta_4$'' model, defined by~\cite{Cusin:2016nzi}
\be\label{D4}
\Gamma_{\Delta_4}=
\frac{\mplr^2}{2}\int d^{4}x \sqrt{-g}\, 
\[R-\frac{m^2}{6} R\frac{1}{\Delta_4} R\]\, ,
\ee
where
\be\label{defDP}
\DP\equiv\Box^2+2\RMN\n_{\mu}\n_{\nu}-\frac{2}{3}R\Box+\frac{1}{3}\gMN\n_{\mu} R\n_{\nu}\, 
\ee
is  the so-called Paneitz operator, which enters in the quantum effective action for the conformal anomaly. 
Conceptual and phenomenological aspects of these models have been discussed in a series of papers 
\cite{Maggiore:2013mea,Foffa:2013sma,Foffa:2013vma,Kehagias:2014sda,Maggiore:2014sia,Nesseris:2014mea,Dirian:2014ara,Cusin:2014zoa,Barreira:2014kra,Dirian:2014xoa,Dirian:2014bma,Mitsou:2015yfa,Barreira:2015vra,Maggiore:2015rma,Cusin:2015rex,Cusin:2016nzi,Dirian:2016puz,Maggiore:2016fbn,Nersisyan:2016hjh}, and recently reviewed in \cite{Maggiore:2016gpx}.

In particular, the study of their cosmological solutions shows that, 
at the background level, they generate  an effective  dark energy and have a realistic background FRW evolution, without the need of introducing a cosmological constant. Their cosmological perturbations  are well-behaved both  in the scalar and in the tensor sector, and give predictions 
consistent with CMB, supernovae, BAO and structure formation 
data.  This allowed us to move to the next step, implementing the cosmological perturbations in a Boltzmann code and performing Bayesian parameter estimation and a detailed quantitative comparison with $\Lambda$CDM~\cite{Dirian:2014bma,Dirian:2016puz}. The result is that the RT model   fits the data at a level which is statistically indistinguishable  from  $\Lambda$CDM (and in fact even fits slightly better). The performance of the RR model is also statistically indistinguishable from that of $\Lambda$CDM, once neutrino masses are left free to vary, within the existing experimental limits~\cite{Dirian:2017pwp,Belgacem:2017cqo}. 
We refer the reader to \cite{Maggiore:2016gpx} and references therein for a detailed discussion of conceptual aspects related to these nonlocal models, as well as  for a discussion of the various nonlocal models that have been studied, and which finally allowed us to narrow down the choice of interesting nonlocal models.\footnote{In particular in \cite{Belgacem:2017cqo}, that was finalized after the first version of this paper, we showed that the $\Delta_4$ model is phenomenologically excluded because it predicts a speed of gravitational waves sensibly different from the speed of light, and therefore inconsistent with GW170817, while the RR and RT models pass this test.}

In the present paper we  address a technical but important point concerning the evolution of these models when the initial conditions are set during a phase of primordial inflation. Indeed, in the phenomenological studies mentioned above, the evolution of the models was started deep in the radiation-dominated (RD) phase. In that case, we found a well behaved background evolution, with stable cosmological perturbations, until the present epoch, for the RT, RR and $\Delta_4$ models. However, these nonlocal models are not low-energy effective theories. We interpret them as quantum effective actions derived from a fundamental action, which could be just the Einstein-Hilbert action; as such, their domain of validity is the same as that of the original theory. Hence, it is natural to demand that a correct model should give a viable and stable cosmological evolution even if we start its evolution during an earlier phase of primordial inflation. We can take this as a further constraint that  could help us to further narrow down the choice of  viable nonlocal models.
We will see in this paper that this requirement is  met by all three models, although for the RT model the issue is more subtle, and there is an apparent instability during a de~Sitter epoch; however further analysis shows that this apparent instability 
is innocuous and also the RT model is viable, even when setting the initial conditions during inflation.\footnote{On this point, we correct the wrong conclusion of the previous version of this paper.}

The paper is organized as follows. In Section~\ref{sect:cosmevol} we study the cosmological evolution of the RT model when the initial conditions are set during inflation, while in Sections~\ref{sect:RR} and \ref{sect:D4} we investigate the same issue in the RR and in the $\Delta_4$ models.

\section{Cosmological evolution of the RT model with initial conditions during inflation} 
\label{sect:cosmevol}

\subsection{Background evolution}\label{sect:RTbackev}

As discussed in \cite{Maggiore:2013mea,Foffa:2013vma,Nesseris:2014mea}, to write the equations of motion of the RT model in  local form one first introduces an  auxiliary field $U\equiv -\iBox R$. Defining 
$S_{\mu\nu}\equiv -U\gmn=\gmn \iBox R $, we can then split this tensor into a  transverse part $S_{\mu\nu}^{\rm T}$, which by definition satisfies $\n^{\mu}S_{\mu\nu}^{\rm T}=0$, and a longitudinal part, 
\be\label{splitSmn}
S_{\mu\nu}=S_{\mu\nu}^{\rm T}+\frac{1}{2}(\n_{\mu}S_{\nu}+\n_{\nu}S_{\mu})\, . 
\ee
Thus, the nonlocal quantity $S_{\mu\nu}^{\rm T}=(\gmn \iBox R)^{\rm T}$ that appears in \eq{RT} can be written as
\be
(\gmn \iBox R)^{\rm T}=-U\gmn-\frac{1}{2}(\n_{\mu}S_{\nu}+\n_{\nu}S_{\mu})\, ,
\ee
in terms of an auxiliary scalar field $U$ and an auxiliary four-vector field $S_{\mu}$. The field $U$  by definition satisfies 
\be\label{BoxUR1}
\Box U=-R\, , 
\ee
 while $S_{\mu}$  satisfies the equation
\be\label{panU}
\Box S_{\nu}+\n^{\mu}\n_{\nu}S_{\mu}=-2\pan U\, ,
\ee
which is obtained by taking the divergence of \eq{splitSmn}. The temporal component $S_0$ of  $S_{\mu}$ is of course a scalar under rotation and, together with $U$, enters the equations at the background level. In contrast, at the background level the spatial vector $S_i$  vanishes because there is no preferred spatial direction in a FRW background. However, $S_i$ enters at the level of cosmological perturbations. In a spacetime such as FRW, which has invariance under spatial rotations, we can split  the vector field $S_i$ into its longitudinal and transverse parts with respect to spatial derivatives, 
$S_i=S^{\rm T}_i+\pa_i S$, where $\pa_iS^{\rm T}_i=0$. Then at the background level $S=0$, but at the level of perturbations $\d S$ is non-vanishing and contributes to the scalar perturbations. In conclusion, at the background level we have two auxiliary fields, $U$ and $S_0$, while for  scalar perturbations the auxiliary fields contribute with  $\d U,\d S_0$ and $\d S$.

Specializing  the $(0,0)$ component of the Einstein equations and \eqs{BoxUR1}{panU} to a FRW metric, one obtains
\bees
H^2-\frac{m^2}{9}(U-\dot{S}_0)&=&\frac{8\pi G}{3}\rho\,\label{loc1} \\
\ddot{U}+3H\dot{U}&=&6\dot{H}+12H^2\, ,\label{loc2}\\
\ddot{S}_0+3H\dot{S}_0-3H^2S_0&=&\dot{U}\, ,\label{loc3}
\ees
where the dot is the derivative with respect to cosmic time $t$.
It is convenient to pass to dimensionless variables,  using
$x\equiv \ln a(t)$ to parametrize the temporal evolution. We denote $df/dx=f'$, and we define $h(x)=H(x)/H_0$, $\zeta(x)\equiv h'/h$, $\g\equiv  m^2/(9H_0^2)$,  and
\be
Y\equiv U-\dot{S}_0\, .
\ee
Then the above equations become
\bees
&&\hspace*{-5mm}h^2(x)=\Omega_M e^{-3x}+\Omega_R e^{-4x}+\g Y(x)\, ,\label{hLCDM}\\
&&\hspace*{-5mm}Y''+(3-\zeta)Y'-3(1+\zeta)Y=3U'-3(1+\zeta)U\, ,\label{sy1}\\
&&\hspace*{-5mm}U''+(3+\zeta)U'=6(2+\zeta)\label{sy3}\, ,
\ees
where  $\Omega_{R,M}\equiv \rho_{R,M}(t_0)/\rho_c$ denotes as usual  the present energy fractions.
We see from \eq{hLCDM} that
there is an effective DE density
$\rde(x)=\rho_0\g Y(x)$,  
where $\rho_0=3H_0^2/(8\pi G)$.

Given the initial conditions of the auxiliary fields, the numerical integration of \eqs{sy1}{sy3} is straightforward. Before proceeding with the study of these equations, it is however important to understand the freedom in the choice of these initial conditions.
It should be stressed that the auxiliary fields, such as $U, S$ and $S_0$  (or $Y$) in the RT model (or $U$ and $S$ in the RR model, see below) do not represent new independent  degrees of freedom of the theory. Their initial conditions are in fact in principle fixed by the initial conditions on the metric. If one had an explicit derivation of the quantum effective action from the fundamental theory, one could derive this relation explicitly. A toy example of a nonlocal quantum effective action that can be derived explicitly from the fundamental theory is the Polyakov quantum effective action in $D=2$ space-time dimensions,
which can be obtained from the fundamental action of two-dimensional gravity coupled to conformal matter fields, through the integration of the conformal anomaly. In the Polyakov quantum effective action the nonlocal term is proportional to $R\iBox R$, and could in principle be written in local form, while still maintaining explicit diffeomorphism invariance, by introducing an auxiliary field $U=-\iBox R$.\footnote{The Polyakov action becomes local also when written in terms of the conformal factor. In this case, however, diffeomorphism invariance is no longer explicit.} In this case, in which one has an explicit derivation of the quantum effective action,  one can also find explicitly the relation between the initial conditions of $U$ and the initial conditions of the metric, see \cite{Maggiore:2016gpx}. This point is conceptually important, since it implies than nonlocal gravity is not the same as a scalar-tensor theory (or scalar-vector-tensor, for the RT model). 

In practice, however, lacking an explicit derivation of the infrared limit of the quantum effective action in the four-dimensional case, one is in principle confronted with the problem of choosing the initial conditions of the auxiliary fields and of their derivatives. At the level of background evolution, these initial conditions are in one-to-one correspondence with the independent solutions of the homogeneous equations associated to the definition of the auxiliary fields, which for the RT model are given by \eqs{loc2}{loc3}.  In general, for such  homogeneous equations, we can have either unstable modes,  growing exponentially with $x$, or exponentially decaying modes, or constant modes. The initial conditions of the auxiliary fields that excite  decaying modes correspond to irrelevant direction in parameter space. Therefore, these parameters do not affect the cosmological predictions of the model, nor its stability, and can be set to zero. Marginal directions can introduce a freedom in the predictions of the theory, but  do not threaten the stability of the model. In contrast, growing modes can pose a threaten, potentially leading to an unacceptable evolution and a strong dependence on the initial conditions, and must be studied with care. Let us recall the results of this analysis for the RT model~\cite{Maggiore:2013mea}. To obtain simple analytic results, it is convenient to  make use of the fact  that, in 
any given epoch,  the parameter $\zeta$ has an approximately constant value $\zeta_0$, with  
$\zeta_0=\{0,-2,-3/2\}$ in de~Sitter (dS),  RD and  MD, respectively. In the approximation of constant $\zeta$ \eq{sy3} can be integrated analytically, and has the solution 
 \be \label{pertU}
U(x)=\frac{6(2+\zeta_0)}{3+\zeta_0}x+u_0
+u_1 e^{-(3+\zeta_0)x}\, ,
\ee
where the coefficients $u_0,u_1$ parametrize the general solution of the homogeneous equation $U''+(3+\zeta_0)U=0$. Plugging \eq{pertU} into \eq{sy1} and solving for $Y(x)$ we get
\bees
&&\hspace*{-5mm}Y(x)=-\frac{2(2+\zeta_0)\zeta_0}{(3+\zeta_0)(1+\zeta_0)}
+\frac{6(2+\zeta_0)}{3+\zeta_0} x+u_0\nn\\
&&\hspace*{-5mm}-\frac{6(2+\zeta_0) u_1 }{2\zeta_0^2+3\zeta_0-3}e^{-(3+\zeta_0)x} 
 +a_1 e^{\a_{+}x}+ a_2 e^{\a_{-}x}\, ,\label{pertY}
\ees
where
\be\label{apm}
2\a_{\pm}=-3+\zeta_0\pm \sqrt{21+6\zeta_0+\zeta_0^2}\, .
\ee
During RD and MD both $\a_{+}$ and $\a_{-}$ are negative. 
This is important, since a positive value would have led to a mode growing exponentially in $x$ (i.e., as a power in the scale factor). Any small perturbation would then unavoidably excite this growing mode, and  this would have quickly led  to an unacceptable background evolution during RD or MD. In contrast, during dS there is a growing mode, with  
\be
\alpha_+^{\rm dS}=(-3+\sqrt{21})/2\simeq 0.79\, .
\ee
This instability is due to the $S_0$ auxiliary field. Indeed, from \eq{loc3} with $H$ constant we see that there is a solution of the associated homogeneous equation $S_0(t)\propto e^{\alpha_+^{\rm dS} Ht}$, i.e.
\be\label{S0ax}
S_0(x)\propto e^{\alpha_+^{\rm dS}  x}\, ,
\ee
which induces the same instability in $Y=U-\dot{S}_0$.
In our previous work on the RT model the cosmological evolution was started deep in the RD phase, with initial values  
$U(x_{\rm in})=U'(x_{\rm in})=Y(x_{\rm in})=Y'(x_{\rm in})=0$. The initial conditions 
$\{U(x_{\rm in}),U'(x_{\rm in}),Y(x_{\rm in}),Y'(x_{\rm in})\}$ are in one-to-one correspondence with the parameters $\{u_0,u_1,a_1,a_2\}$ in \eqs{pertU}{pertY}. From the fact that, in RD, $\a_{\pm}<0$, it follows that $a_1,a_2$ correspond to stable directions, i.e. the solution obtained starting in RD with generic non-vanishing values of $a_1,a_2$ approaches exponentially fast the solution obtained starting with 
$a_1=a_2=0$. Similarly, we see from \eq{pertU} (with $\zeta_0=-2$ in RD) that $u_1$ is associated to an irrelevant direction. In contrast, $u_0$ is associated with a marginally-stable direction, and  setting a non-vanishing  initial value for $u_0$ effectively amounts to reintroducing a cosmological 
constant~\cite{Maggiore:2013mea,Foffa:2013vma}. 

Let us now investigate what happens 
if we start the evolution  at some initial time deep into a primordial de~Sitter   inflationary phase with generic initial conditions, i.e. without fine-tuning $Y$ to zero at the initial time. At first sight, one might fear that the presence of the unstable mode (\ref{S0ax}) should lead to an unacceptable evolution and to a strong dependence on the associated initial conditions. However, we will see that the situation is acceptable, both at the level of background evolution and of cosmological perturbations.

We denote by $a_{\rm in}$ the value of the scale factor, during inflation, when we set the initial conditions, and by $a_{\rm end}$ the value  when inflation ends. For the purpose of studying the stability of the model, we will neglect the intermediate reheating phase (which in typical reheating models is similar to a matter-dominated phase).
We write  
\be
x_{\rm end}-x_{\rm in}=\log \(a_{\rm end}/a_{\rm in}\)\equiv \Delta N\, ,
\ee 
i.e. $\Delta N$ is the number of e-folds from the initial time to the end of a quasi-de~Sitter phase of inflation. Thus, if $Y(x_{\rm in})$ has a generic value of order one (i.e., is not fine-tuned to zero), by the end of inflation 
\be
Y(x_{\rm end})\simeq \exp\{\alpha_+^{\rm dS} \Delta N\}\simeq \exp\{0.79\Delta N\}\, . 
\ee
Note that, despite this exponential growth, even for very large values of  $\Delta N$  the corresponding  DE density $\rde(x)=\rho_0\g Y(x)$  has no effect on the inflationary dynamics. This is due to the fact that $\rho_0=3H_0^2/(8\pi G)\sim (10^{-3} {\rm eV})^4$ is extremely small compared to the energy density during inflation. For instance, if  $Y(x_{\rm in})={\cal O}(1)$ and we take $\Delta N=60$, at the end of inflation we get $Y(x_{\rm end})={\cal O}(10^{20})$.
Even with such a large value of $Y$, we have
\be
[\rho_0 Y(x_{\rm end})]^{1/4}\sim 10^{-3} {\rm eV}\times Y^{1/4}(x_{\rm end})={\cal O}(10^2)
\, {\rm eV}\, .
\ee   
This is totally negligible  compared to the inflationary scale  $M$, that has  typical values, say, of order $10^{13}$~GeV.\footnote{Furthermore,  we will see below that  $\gamma\ll 1$, and decreases  with $\Delta N$ so to keep
 $\gamma Y(x=0)$ fixed. } Thus, during the inflationary phase the evolution of the scale factor is the same  as in standard GR without the nonlocal term; the auxiliary fields $U$ and $Y$ are at this stage `spectator fields', that do not influence the evolution of the scale factor, and  whose evolution can be computed inserting in \eqs{sy1}{sy3} the  solution for $H(t)$ computed from the inflationary solution, neglecting the nonlocal term. 
Observe that \eqs{sy1}{sy3} are not affected by the inclusion of the  inflationary sector, since they just express the definition of the auxiliary fields (e.g $U=-\iBox R$), needed to put the original nonlocal equations in local form.

The evolution of $U$ can be computed similarly, using \eq{pertU}. During a quasi-de~Sitter phase of inflation, starting from a value of order one, we get 
\be\label{Uxend}
U(x_{\rm end})\simeq 4\Delta N\, . 
\ee
We now  use \eqst{hLCDM}{sy3} to further evolve numerically the system through RD, MD and the present DE-dominated epoch, using $Y(x_{\rm end})$ and $U(x_{\rm end})$ as initial values for the subsequent evolution.

\begin{figure}[t]
\includegraphics[width=0.45\textwidth]{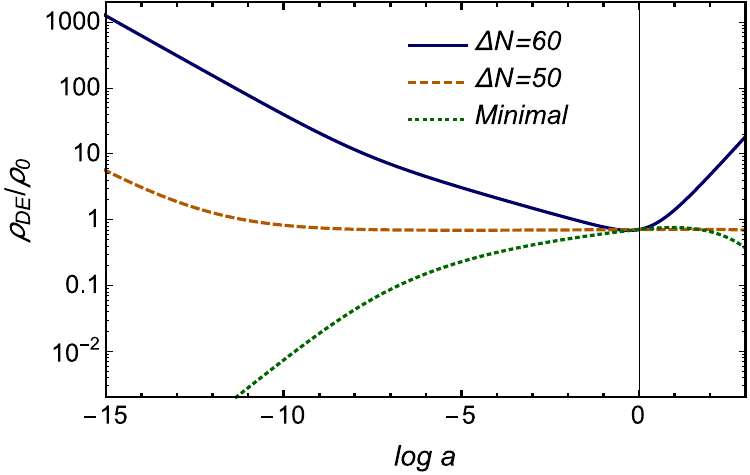}
\includegraphics[width=0.45\textwidth]{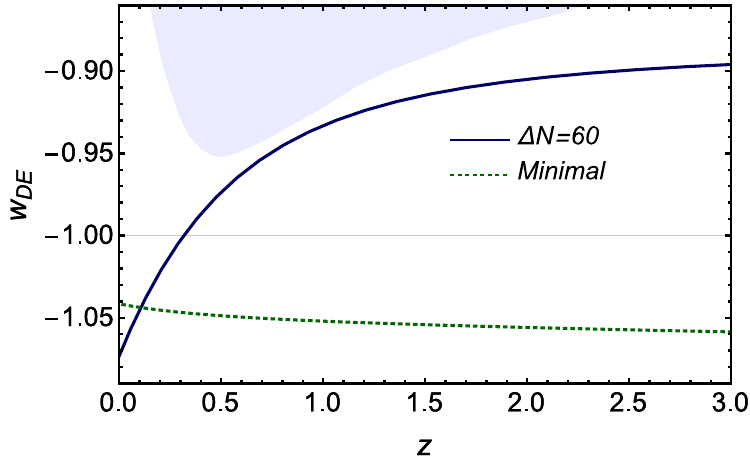}
\caption{Left panel: the evolution of the dark energy density for  $\Delta N=50$ and $\Delta N=60$, compared to  the ``minimal" scenario  where the evolution is started in RD with vanishing initial conditions on the auxiliary fields. Right panel: The prediction for  $w_{\rm DE}(z)$ for $\Delta N\,\gsim 60$. The shaded region is excluded by the Planck data~\cite{Ade:2015rim}.}
\label{fig:rho_vs_x}
\end{figure}

The results are shown in Fig.~\ref{fig:rho_vs_x}. In the left panel 
we show the result for the DE density $\rde(x)$ as a function of $x=\log a$, for $\Delta N=50$ and $\Delta N=60$, compared to the `minimal scenario' where the evolution is started in RD with vanishing initial conditions. For each value of $\Delta N$, the parameter $\gamma$ is chosen in such a way to obtain the observed present value of  $\rde(x=0)/\rho_0$, that here we have fixed to $0.7$ (of course, in a full analysis including the cosmological perturbations, this value will be determined self-consistently by Bayesian parameter estimation). Explicitly, we find $\gamma\simeq\{0.05,0.005,3\times 10^{-4}\}$
for the minimal model and for $\Delta N=50,60$, respectively.\footnote{For sufficiently large $\Delta N$, an increase in the initial values of $Y$ at the beginning of RD is exactly compensated by a decrease in $\gamma$, and we end up on the same solution. This is due to the fact that, for large $\Delta N$, after inflation $U\ll Y$ and we can set to zero the right-hand side in \eq{sy1}. Then  we get a homogeneous equation for $Y$ and, for the subsequent evolution, an increase in the  values $Y(x_{\rm end})$ and
$Y'(x_{\rm end})\simeq \a_{+}^{\rm dS}Y(x_{\rm end})$ can be exactly compensated by a decrease in $\gamma$, such that  $\rde\equiv \rho_0\gamma Y$ is unchanged.\label{foot:gDN}}

Observe that RD-MD equilibrium is at $x_{\rm eq}\simeq -8.1$, while $x=0$ corresponds to the present epoch, and we extended the plot into the cosmological future. 
The right panel shows the corresponding result for the DE equation of state $\wde(z)$, defined 
as usual from the conservation equation
\be
\dot{\rho}_{\rm DE}+3(1+w_{\rm DE})H\rho_{\rm DE}=0\, , 
\ee
We see that, at the background level, we get a viable and in fact even quite interesting evolution, with a DE equation of state that, for $\Delta N\, \gsim 60$, crosses from non-phantom to phantom with decreasing redshift.

\subsection{Cosmological perturbations}

We next study what happens at the level of cosmological perturbations. The equations governing the cosmological perturbations of the RT model have been written down  in \cite{Nesseris:2014mea} and in
app.~A of 
\cite{Dirian:2014ara}. In the scalar sector, the metric perturbations are written as 
\be\label{defPhiPsi}
ds^2 =  -(1+2 \Psi) dt^2 + a^2(t) (1 + 2 \Phi) \delta_{ij} dx^i dx^j\, ,
\ee
We further expand the auxiliary fields into a background value plus perturbations. 
As discussed above, 
at the perturbation level in the scalar sector we have three variables
$\d U,\d S_0$ and $\d S$.  It is convenient to trade $S_0$ and $S$ for the variables\be\label{defVZS0S}
V=H_0S_0\, ,\qquad
Z=H_0^2 S\, .
\ee
where $S_0$ is the $\mu=0$ component of $S_{\mu}$ in coordinates $(t,\vx)$.\footnote{Observe that, in  app.~A of ref.~\cite{Dirian:2014ara}, we defined $V$ as $V=H_0S_0/a$, where $S_0$ 
was defined as the $\mu=0$ component of $S_{\mu}$ in coordinates $(\eta,\vx)$, where $\eta$ is conformal time. Denoting by $S_t$ the $\mu=0$ component of $S_{\mu}$ in coordinates $(t,\vx)$ and by $S_{\eta}$ the $\mu=0$ component of $S_{\mu}$ in coordinates $(\eta,\vx)$, we have $S_t dt=S_{\eta} d\eta$ and therefore $S_{\eta}=aS_t$. In this paper we rather use the notation $S_0$ for  the $\mu=0$ component of $S_{\mu}$ in coordinates $(t,\vx)$.} 
We then  use
$\d U, \d V, \d Z$. The corresponding perturbed Einstein equations are
\bees
&&\hspace{-8mm}\hat{k}^2 \Phi + 3(  \Phi' - \Psi) = \frac{3}{2h^2\rho_0}
\[ \delta \rho + \gamma \rho_0 \( \delta U - h  \delta V' + 2 h \Psi \bar{V}' + h  \Psi' \bar{V} \) \], \label{EE1xRT}\\
&&\hspace{-8mm} \hat{k}^2 (\Phi' - \Psi ) = - \frac{3}{2h^2\rho_0}
\[ \bar{\rho} (1+w) \hat{\theta} + \hat{k}^2 \gamma \rho_0 
\(  h^2 \delta Z - \frac{h^2}{2} \delta Z' + h \Psi \bar{V} - \frac{h}{2} \delta V \) \], \label{EE2xRT} \\
&&\hspace{-8mm} \hat{k}^2 (\Psi + \Phi-3\gamma\d Z) = \frac{9 }{2h^2\rho_0} 
\bar{\rho} (1+w) e^{2x}\sigma\, ,  \label{EE3xRT}\\
&& \hspace{-8mm} \Phi'' + (3+\zeta)  \Phi' -  \Psi' - (3+2\zeta) \Psi + \frac{\hat{k}^2}{3} (\Phi + \Psi) \nonumber \\
&& = - \frac{3}{2h^2\rho_0}
\[ \delta p - \gamma \rho_0 \( \delta U - h (  \Phi' - 2 \Psi ) \bar{V} - h \delta V - \frac{\hat{k}^2}{3} h^2 \delta Z \)  \]\, , \label{EE4xRT}
\ees
where $\hat{k}=k/(aH)$,  $\hat{\theta}=\theta/(aH)$, an overbar denotes here a background quantity
and, as before, the prime denotes the derivative with respect to $x\equiv \ln a$, and $\zeta(x)\equiv h'/h$.
The linearization of the equations for the auxiliary fields  gives 
\bees
&& \hspace{-10mm} \delta U'' + (3 + \zeta)  \delta U' + \hat{k}^2 \delta U = 2 \hat{k}^2 (\Psi + 2 \Phi)   + 6 ( \Phi'' +(4+\zeta)  \Phi' ) - 6 \[ \Psi' + 2(2+ \zeta) \Psi \] \nonumber \\
&& \hspace{5mm}
+ 2 \Psi  \bar{U}'' +
\[ 2 \Psi (3+\zeta) + (\Psi' - 3  \Phi') \]  \bar{U}'\,\label{AppU} \\
&&\hspace{-10mm} \delta V'' + (3+\zeta)  \delta V' 
+ \frac{\hat{k}^2}{2} h (  \delta Z' - 4 \delta Z )- h^{-1} \d U'  =  2 \Psi  \bar{V}''
 + \[ 2(3+\zeta) \Psi + 3( \Psi' -  \Phi') \]  \bar{V}' \nonumber \\
&&  \hspace{5mm}
 + \[ \Psi'' + (3 + \zeta)  \Psi' + 6 \Phi' \] \bar{V}  - \[ (1/2) \hat{k}^2 - 3 \] \( \delta V - 2 \Psi \bar{V} \)\, ,
\label{AppV} \\
&& \hspace{-10mm}
 \delta Z'' + (1+\zeta)\delta Z' + 2 \(  \hat{k}^2 - (3+\zeta) \) \delta Z =2h^{-2}\d U\nn\\
&&\hspace{5mm}-h^{-1}\[  \delta V' + 5 \delta V - 4 \Psi  \bar{V}' - 2 (  \Psi' -  \Phi' + 4 \Psi ) \bar{V} \]
\label{AppZ}\, .
\ees
These equations have been studied in RD and in MD in \cite{Nesseris:2014mea,Dirian:2014ara}.
Let us now see  what happens during a phase of primordial inflation. 
By construction, a successful model of primordial inflation must be such that, during the inflationary phase, 
all the modes relevant for cosmology eventually exit the horizon.  It can be natural to set initial conditions on the auxiliary fields at the time that they exit the horizon (see the discussion below for the proper initial conditions), so we need to study
the stability  of the perturbation equations for super-horizon modes. In this regime the physical wavelength is much larger  than $H^{-1}$ and the physical momentum $k/a\ll H $, so
$\hat{k}\ll 1$.  Then,
in the above equations we  keep only the lowest-order terms in $\hat{k}^2$.
We also limit ourselves for simplicity to an inflationary  phase in the limit of a de~Sitter expansion, supported by a constant vacuum energy density. Then the matter perturbations $\d\rho,\hat{\theta}$ and $\sigma$ vanish in the exact de~Sitter limit.

Furthermore
the dimensionless Hubble parameter $h(x)=H(x)/H_0$ becomes constant and is numerically very large for all typical inflationary scales.  Indeed,
\be\label{eq1}
h_{\rm infl}\equiv \frac{H_{\rm infl}}{H_0}=\(\frac{\rho_{\rm infl}}{\rho_0}\)^{1/2}\simeq
\(\frac{M_{\rm infl}}{3\times 10^{-3} {\rm eV}}\)^2\, ,
\ee
where  $\rho_{\rm infl}$ is the energy density during a phase of de~Sitter inflation, and $M_{\rm infl}\equiv \rho_{\rm infl}^{1/4}$ is the corresponding mass scale. We can then show that all terms in the above equation involving $\bar{V}(x)$ can be set to zero. Indeed, from $Y=U-\dot{S}_0$, it follows that $V'=(U-Y)/h$. Using $\bar{U}(x)=4x$ and $\bar{Y}(x)\simeq \exp\{0.79 x\}$ (apart from an overall constant of order one) it follows that, during inflation,
$\bar{V}(x)={\cal O}(h_{\rm infl}^{-1} \exp\{0.79 x\})$. Thus  $\bar{V}$ reaches a maximum value at the end of inflation of order $h_{\rm infl}^{-1} \exp\{0.79 \Delta N\}$. Despite the exponential term, this is still an extremely small number, because $h_{\rm infl}$ is huge. For instance, for
$M_{\rm infl}= 10^{13}\, {\rm GeV}$, from \eq{eq1}  we have $h_{\rm infl}\sim 10^{49}$ and, even setting $\Delta N=60$, $h_{\rm infl}^{-1} \exp\{0.79 \Delta N\}\sim 10^{-49}\times 10^{20}$ is totally negligible. More generally, 
for inflation at a scale $M_{\rm infl}$, the minimum number of efolds necessary for solving the horizon and flatness problems is given by 
\be
\Delta N\simeq 64-\log\( \frac{10^{16}\, {\rm GeV} }{M_{\rm infl}} \)\, ,
\ee
see e.g. Sect.~21.1 of \cite{Maggiore:2018zz} (this expression neglects the effect of a phase of reheating, which is adequate here given the orders of magnitude involved). Combining this with \eq{eq1} we get 
\be\label{hin2DeltaN}
h_{\rm infl}^{-1} e^{0.79 \Delta N} \simeq 10^{-33}\, \( \frac{10^{16}\, {\rm GeV} }{M_{\rm infl}} \)^{1.21}\, ,
\ee
that even for $M_{\rm infl}$ as low as 1~TeV is at most of order $10^{-18}$. Since all occurrences of $\bar{V}$ in the perturbation equations appear multiplied overall by factors ${\cal O}(1)$ or even ${\cal O}(1/h)$,
we can set $\bar{V}(x)=0$ in the above equations. In contrast,   from
\eq{pertU}, $\bar{U}'=4$ is a number $O(1)$.

As usual, because of general covariance, in the scalar sector the four  linearized Einstein equations (\ref{EE1xRT})--(\ref{EE4xRT}) are not independent once we take into account the linearized energy-momentum conservation, and we have only two independent equations. We take them to be 
\eqs{EE1xRT}{EE3xRT} which, together with \eqst{AppU}{AppZ} give a system of five equations for the five functions $\Psi,\Phi,\d U,\d V$ and $\d Z$.

To take the limit $\hat{k}^2\ll 1$  in
\eq{EE3xRT} we observe that, on the left-hand side, we have $\hat{k}^2=k^2/(a^2H^2)$, while on the right-hand side we have $\bar{\rho}e^{2x}/(h^2\rho_0)$. Since $e^{2x}=a^2$ and $\bar{\rho}\sim \rho_0/a^4$, this term is of order $H_0^2/(a^2 H^2)$. All modes of cosmological interest are inside the horizon at the present time, so they have $k>H_0$.\footnote{Recall that we are using units such that the scale factor today is equal to one, so the comoving momontum $k$ is the same as the physical momentum today.} Thus, even if we take the limit $\hat{k}\ra 0$, the factor $\hat{k}^2$ on the left-hand side of \eq{EE3xRT} is larger than the factor $\bar{\rho}e^{2x}/h^2\rho_0$ on the right-hand side. On top of this,
on the right-hand side we also have the factor $(1+w) e^{2x}\sigma$, which 
vanishes in an exact de~Sitter phase.\footnote{More precisely, if,  rather than an exact de~Sitter phase, we consider slow-roll inflation sourced by a single inflation field $\phi$, the anisotropic stress tensor 
 $\Sigma_{ij}$ is given by 
$\pa_i\d\phi\pa_j\d\phi$. The definition of $\sigma$ in terms of $\Sigma_{ij}$ is
$\bar{\rho} (1+w) \sigma \equiv a^{-2}\n^{-2} \partial^i \partial^j \Sigma_{ij}$
 (we follow the notation in \cite{Dirian:2014ara}).
It then follows that
$\bar{\rho} (1+w) \sigma$ is  of second order in the perturbations, and when working at linear order it can be neglected.} Thus, in this limit \eq{EE3xRT} becomes
$\Psi + \Phi-3\gamma\d Z=0$.

 It is  convenient to introduce rescaled variables $u,v,z$ from $\d U=h^2u$, 
$\d V=h v$ and, for uniformity of notation, $\d Z=z$. Then the independent equations take the following form, for super-horizon modes during inflation:
\bees
\Phi'-\Psi&=&\frac{1}{2}\gamma (u-v')\, ,\label{eqrescaled1}\\
\Psi + \Phi&=&3\gamma z\, ,\label{eqrescaled1b}\\
u''+3u'&=&(6\Phi''+12\Phi'-2\Psi')/h^2\, ,\label{eqrescaled2}\\
v''+3v'-3v&=&u'\, ,\label{eqrescaled3}\\
z''+z'-6z&=&2u-v'-5v\, .\label{eqrescaled4}
\ees
We observe that, with this rescaling, $h$ disappears from \eq{eqrescaled1}. Furthermore, in the limit $h\ra\infty$, the right-hand side of \eq{eqrescaled2} vanishes and the equations for the perturbations of the auxiliary fields, $u,v$ and $z$, decouple from the matter perturbations. Solving the homogeneous equations associated with \eqst{eqrescaled2}{eqrescaled4} we see that $z$ has a mode growing as $e^{2x}$ and $v$  a mode growing as $e^{\alpha_{+}^{\rm dS}x}\simeq e^{0.79x}$.  By the end of inflation, within this linearized approximation, $z$ has grown by a factor $e^{2\Delta N}$ compared to its initial value, while $v$ by a factor $e^{0.79\Delta N}$.
Taking for instance $\Delta N=60$, we have
$e^{2\Delta N}\sim 10^{52}$. 
At first sight, 
this leads  to the conclusion that these perturbation variables leave the linear regime. If this were the case then, through their coupling with the metric perturbations in \eq{eqrescaled1}, they would spoil cosmological perturbation theory and the initial conditions $\Phi,\Psi\sim 10^{-5}$ provided by inflation in the standard scenario, leading to an unacceptable cosmology.

However, this conclusion is incorrect because it does not take into account that, because of the rescaling performed, the natural values of the initial conditions on the variables $u,v$ and $z$ are extremely small. In terms of the original auxiliary field $U\equiv -\iBox R$, a natural initial condition, say set at the time that the modes exit the horizon during inflation, could be $U_{\rm in}={\cal O}(1)$.\footnote{The precise value will be of no importance; we will see from the final result that it is enough that the initial conditions set at this time are not astronomically large numbers, say $10^{55}$, see \eq{hin2DeltaN} below.} However $u(t)\equiv U(t)/h(t)^2$, where $h(t)=H(t)/H_0$. As we have seen in \eq{eq1}, 
during a phase of primordial inflation $h^2(t)\equiv h_{\rm infl}^2$ is a huge number. For instance, for
$M_{\rm infl}= 10^{13}\, {\rm GeV}$,   we have $h^2_{\rm infl}\sim 10^{98}$. Thus, the natural initial condition on $u$ during inflation is 
\be
u_{\rm in}=h^{-2}_{\rm infl}U_{\rm in}\, ,
\ee
which, for $M_{\rm infl}= 10^{13}\, {\rm GeV}$, means $u_{\rm in}\sim 10^{-98}U_{\rm in}$. The same happens for the variables $v$ and $z$ (which are those that display instabilities). From the $\mu=\nu=0$ component of eq.~(2.2) we see that, using cosmic time, so that $\gmn=(-1, a^2\delta_{ij})$, the natural initial condition on $S_0$ is such that $(\pa_0 S_0)_{\rm in}\sim U_{\rm in}$ . Using  $\pa_0\sim H$ we get $(S_0)_{\rm in}\sim H_{\rm infl}^{-1}U_{\rm in}$. Therefore, for the variable
$V=H_0S_0$, we have $V_{\rm in}\sim U_{\rm in} H_0/H_{\rm infl}=h_{\rm infl}^{-1} U_{\rm in}$.
The rescaled variable $v$ is  defined by 
$v=V/h$ and  therefore 
\be
v_{\rm in}\sim h_{\rm infl}^{-2} U_{\rm in}\, .
\ee
Similarly, from the $(ii)$ component of eq.~(2.2) and the definition of $S$ from $S_i=S^{\rm T}_i+\pa_i S$, we see that the natural initial conditions on the momentum modes $S_{\bf k}$ are given by
$Ua^2\sim k^2S_{\bf k}$, where $k$ is the comoving momentum. As we mentioned, a natural time  to impose initial conditions is when the mode is crossing the horizon, so that $k^2/a^2\sim H_{\rm infl}^2$. Then 
$S_{\rm in}\sim U_{\rm in}/H_{\rm infl}^2$ and for $z\equiv H_0^2 S$ we get   initial conditions 
\be
z_{\rm in}\sim h_{\rm infl}^{-2} U_{\rm in}\, ,
\ee 
just as for $u$ and $v$. 
The mode that grows faster is $z$, which during de~Sitter grows as $e^{2x}$. Thus, starting from an initial value $\sim h_{\rm infl}^{-2}$, it reaches a value at the end of inflation of order 
$h_{\rm infl}^{-2} e^{2\Delta N}$. Using \eq{eq1}, we see that this is still an extremely small value,
\be\label{hin2DeltaN}
h_{\rm infl}^{-2} e^{2\Delta N} \simeq 10^{-55}\, \( \frac{10^{16}\, {\rm GeV} }{M_{\rm infl}} \)^2\, ,
\ee
that for $M_{\rm infl}= 10^{13}\, {\rm GeV}$ is of order $10^{-49}$ and 
even for $M_{\rm infl}$ as low as 1~TeV is still of order $10^{-29}$. Thus, the perturbations of the auxiliary fields always remain minuscule during de~Sitter inflation, despite their formally growing modes.

A posteriori, the physical interpretation of this result is obvious. The energy scale associated to the nonlocal term is so small compared to the inflationary scale that the nonlocal term has no impact whatsoever on the evolution during inflation. This is  precisely what we found in Sect.~\ref{sect:RTbackev} for the background evolution, where again we found an instability in the variable $Y$, that however has no consequence on the dynamics of the scale factor  during inflation. 

The RT model therefore has a viable cosmological evolution, both at the background level and at the level of perturbations, during all cosmological phases, despite the presence of  an instability during an early inflationary phase. The basic reason is that  the energy scale associated to the nonlocal term is totally negligible with respect to the inflationary scale.

One might wonder what happens in the present cosmological epoch, where, in the RT model, we are approaching a dark-energy dominated phase driven by the nonlocal term, so that now the energy density associated to the nonlocal term is precisely of the same order as the total energy density.  Indeed, in this case the numerical integration shows that the cosmological perturbations eventually become unstable and leave the linear regime, but this happens only in the cosmological future, see Fig.~4 of \cite{Maggiore:2016gpx}.

\section{Stability of the RR model}\label{sect:RR}

We now ask the same question for the RR model. The background evolution has been studied  in
\cite{Maggiore:2014sia}. The background equations can be written in terms of two auxiliary fields $U,V$ as
\bees
&&h^2(x)=\frac{\Omega_M e^{-3x}+\Omega_R e^{-4x}+(\g/4) U^2}
{1+\g[-3V'-3V+(1/2)V'U']}\,,\label{syh2}\\
&&U''+(3+\zeta) U'=6(2+\zeta)\, ,\label{syU2}\\
&&V''+(3+\zeta) V'=h^{-2} U\label{syV}\, .
\ees
where again $\gamma= m^2/(9H_0^2)$ and $\zeta=h'/h$.
The equation for $U$ is the same as in the RT model, so in any given phase with $\zeta=\zeta_0$ the  homogeneous solution for $U$ is again  
\be
U_{\rm hom}(x)=u_0+u_1 e^{-(3+\zeta_0)x}\, . 
\ee
Similarly, the homogeneous equation for $V$ is the same as that for $U$, so 
$V_{\rm hom}(x)=v_0+v_1 e^{-(3+\zeta_0)x}$. In the early Universe we have $-2\leq \zeta_0\leq 0$ and all these terms are either constant or exponentially decreasing, which means that the 
solutions for both $U$ and $V$ are stable  in MD, RD, as well as in a previous  inflationary stage. In particular, in a de~Sitter inflationary epoch $\zeta_0= 0$ and we have a constant mode and a mode decreasing as $e^{-3x}$, for both $U$ and $V$. At the background level, it can be convenient to trade $V$ for another auxiliary field $W(x)=h^2(x)V(x)$. Then \eqst{syh2}{syV} take a form similar to 
\eqst{hLCDM}{sy3}, namely
\bees
&&h^2(x)=\Omega_M e^{-3x}+\Omega_R e^{-4x}+\g Y\label{syh}\\
&&U''+(3+\zeta) U'=6(2+\zeta)\, ,\label{syU}\\
&&W''+3(1-\zeta) W'-2(\zeta'+3\zeta-\zeta^2)W= U\, ,\label{syW}
\ees
where 
\be\label{defYRR}
Y\equiv \frac{1}{2}W'(6-U') +W (3-6\zeta+\zeta U')+\frac{1}{4}U^2\, .
\ee
In this form we see  more easily that  there is again an effective dark energy density, given by $\rde=\rho_0\gamma Y$. The modes associated with the homogeneous equation for $W$ are 
$W_{\rm hom}(x)=w_1e^{-(3-\zeta_0)x}+w_2 e^{2\zeta_0x}$, which are constant or exponentially decreasing in all cosmological epochs. 

The background evolution during a phase of de~Sitter inflation has been studied in section~7.4.1 of \cite{Maggiore:2016gpx}. For $U$ the evolution equation is the same as in the RT model, so at the end of inflation we still have the result (\ref{Uxend}). In contrast, $V$ remains ${\cal O}(1)$  at the end of inflation. Since $V$ is exponentially decreasing during RD and MD, as far as $V$ is concerned the solution obtained starting with $V(x_{\rm end})={\cal O}(1)$ is indistinguishable from that obtained setting
$V(x_{\rm end})=0$. Thus, at the background level, the only  parameter in the initial condition which needs to be taken into account is $u_0$. As shown in \cite{Maggiore:2016gpx}, a large value of $u_0$ (say, $u_0\simeq 240$, corresponding to \eq{Uxend} with $\Delta N\simeq 60$) brings the evolution of the model closer and closer to that of $\Lambda$CDM, see in particular Fig.~10 of \cite{Maggiore:2016gpx}.

\subsection{Cosmological perturbations}

The cosmological perturbations of the RR model during RD, MD and through the present DE-dominated epoch has been studied in detail in \cite{Dirian:2014ara}, where it has been found that they are stable, and well compatible with the cosmological observations.
Here we investigate the stability of the cosmological perturbations of the RR model during a phase of primordial inflation, similarly to what we have done in the previous section for the RT model.
The perturbed Einstein equations are   now~\cite{Dirian:2014ara}
\bees
&&\hspace*{-8mm}\( 1 - 3 \gamma \bar{V} \) \( \hat{k}^2 \Phi + 3  \Phi' -3 \Psi  \) + \frac{3 \gamma}{2} \bigg[  - \frac{1}{2 h^2} \bar{U} \delta U + \big( 6 \Psi - 3 \Phi' - \Psi \bar{U}' \big)  \bar{V}' 
 \nn \\
&& + \frac{1}{2} \big( \bar{U}'  \delta V' +  \bar{V}'  \delta U' \big) - 3 \delta V  - 3 \delta V' - \hat{k}^2 \delta V \bigg] = \frac{3}{2 \rho_0 h^2} \delta\rho\, .\label{eqlin00}\\
&&\hspace*{-8mm}\( 1 - 3 \gamma \bar{V} \)  \hat{k}^2 (  \Phi' - \Psi) - \frac{3 \gamma \hat{k}^2}{2}
\[ \delta V' -  \bar{V}' \Psi - \delta V + \frac{1}{2} \(  \bar{U}' \delta V +\bar{V}' \delta U \)  \] \nn\\
&&= - \frac{3}{2 \rho_0 h^2}  \hat{\theta} \bar{\rho} ( 1 + w)\, ,\label{eqlin0i}\\
&&\hspace*{-8mm}
( 1 - 3 \gamma \bar{V}) \hat{k}^2 (\Psi + \Phi)  - 3 \gamma \hat{k}^2 \delta V = \frac{9}{2 \rho_0 h^2}  e^{2x} \bar{\rho} ( 1 + w) \sigma\, ,\label{eqlintransvij}\\
&&\hspace*{-8mm} ( 1 - 3 \gamma \bar{V} )\[   \Phi'' + (3+\zeta) \Phi' -  \Psi' - (3+2 \zeta) \Psi + \frac{\hat{k}^2}{3}(\Phi + \Psi) \] 
 \nn \\ 
&& - \frac{3 \gamma}{2} \bigg[ \frac{1}{2 h^2} \bar{U} \delta U - 2 \Psi  \bar{V}'' + 
\big[ 2 \Phi' - 2 (2 + \zeta) \Psi -  \Psi' - \Psi  \bar{U}' \big]  \bar{V}' +  \delta V'' + (2+\zeta)  \delta V' \nn \\
&& \hspace{1cm} + \frac{2 \hat{k}^2}{3} \delta V + ( 3 + 2 \zeta ) \delta V + \frac{1}{2} \big(  \bar{U}'  \delta V'  +  \bar{V}'   \delta U' \big) \bigg] = - \frac{3}{2 \rho_0 h^2}  \delta p\, ,
\label{eqlintraceij}
\ees
where we have again used overbars for the background auxiliary fields.
Even at the level of perturbations there are now only two auxiliary fields, and the linearization of their equations gives 
\bees
&& \hspace{-10mm}\delta U'' + (3 + \zeta)  \delta U' + \hat{k}^2 \delta U - 2 \Psi \bar{U}'' - \big[ 2 (3+\zeta) \Psi  +  \Psi' - 3  \Phi' \big]  \bar{U}'  \nn\\
&&  = 2 \hat{k}^2 (\Psi + 2 \Phi) + 6 \big[\Phi'' +(4+\zeta)  \Phi' \big] 
- 6 \big[ \Psi' + 2(2+ \zeta) \Psi \big], \label{eqlinU}\\
&& \hspace{-10mm}\delta V'' + (3 + \zeta) \delta V' + \hat{k}^2 \delta V - 2 \Psi  \bar{V}'' - 
\[ 2 (3+\zeta) \Psi  +  \Psi' - 3  \Phi' \]  \bar{V}' = h^{-2} \delta U\, .\label{eqlinV}
\ees
We now proceed as in the RT model, setting $\zeta=0$ during de~Sitter. We consider first  the modes with $\hat{k}\ll 1$.  
We take again \eqs{eqlin00}{eqlintransvij} as the independent Einstein equations, that, together with
\eqs{eqlinU}{eqlinV}, give four equations for the four functions $\Phi,\Psi,\d U$ and $\d V$.

We also use  the background results $\bar{U}'=4$ and $\bar{V}(x)=0$ and we observe that, since $\bar{U}=O(1)$ and $h\gg 1$ during primordial inflation, we can neglect the terms involving 
$\bar{U}/h^2$.
By writing $\d U=u$ and $\d V=v$ in order to use the same notation that we adopted in the RT model (in the RR case it is not necessary to rescale variables), the independent perturbed equations can now be written as
\bees
\Phi'-\Psi&=&\frac{1}{2}\gamma (v'+3v)\, ,\label{eqRRrescaled1}\\
\Phi + \Psi&=&3\gamma v\, ,\label{eqRRrescaled1b}\\
u''+3u'&=&6\Phi''+12\Phi'-2\Psi'\, ,\label{eqRRrescaled2}\\
v''+3v'&=&0\, .\label{eqRRrescaled3}
\ees
Both \eq{eqRRrescaled3} and the  homogeneous equation $u''+3u'=0$ associated with \eq{eqRRrescaled2}  have only constant or exponentially decreasing solutions for $v$ and $u$, respectively. This means that  there is no instability in the evolution of these variables.

Actually, 
the above analysis   can be easily generalized  to generic values of $\hat{k}$.  In the de~Sitter limit, 
the four independent equations become
\bees
\Phi'-\Psi+\frac{\hat{k}^2}{3}\Phi&=&\frac{1}{2}\gamma (v'+3v)\, ,\label{eqRRkrescaled1}\\
\Phi + \Psi&=&3\gamma v\, ,\\
u''+3u'+\hat{k}^2 u&=&6\Phi''+12\Phi'-2\Psi'+2\hat{k}^2(\Psi+2\Phi)\, ,\label{eqRRkrescaled2}\\
v''+3v'+\hat{k}^2 v&=&0\, .\label{eqRRkrescaled3}
\ees
For every value of $\hat{k}$, the homogeneous equations associated with \eqs{eqRRkrescaled2}{eqRRkrescaled3} still have constant or exponentially decreasing solutions. Indeed they have the form  $e^{\alpha x}$, where $\alpha$ is a (generally complex) solution of the equation $\alpha^2+3 \alpha +\hat{k}^2=0$, and it is easy to see that $\operatorname{Re}(\alpha)\leq 0$.

We conclude that, in the RR model,  cosmological perturbations are stable not only during the  RD and MD phases, but also during a phase of primordial inflation.

\section{Stability of the $\Delta_4$ model}\label{sect:D4}

We finally consider the $\Delta_4$ model (\ref{D4}).
The evolution of a FRW background for the $\Delta_4$ model has been discussed in \cite{Cusin:2016nzi}. The resulting equations, written in terms of two auxiliary fields $U$ and $V$, are
\bees
&&h^2(x)=\frac{\Omega_M e^{-3x}+\Omega_R e^{-4x}+(\g/4) U^2}
{1+\g[-3V'-3V+(1/2)V'U'+V'U]}\,,\label{syh2D4}\\
&&U''+(5+\zeta) U'+2(3+\zeta) U=6(2+\zeta)\, ,\label{syUD4a}\\
&&V''+(1+\zeta) V'=h^{-2} U\label{syVD4}\, ,
\ees
where $\gamma= m^2/(9H_0^2)$ and $\zeta=h'/h$, as in the previous models.
As in the RR model, at the background level, it is convenient to trade $V$ for another auxiliary field $W(x)=h^2(x)V(x)$. Then \eqst{syh2D4}{syVD4} can be written as
\bees
&&h^2(x)=\Omega_M e^{-3x}+\Omega_R e^{-4x}+\g Y\label{syhD4}\\
&&U''+(5+\zeta) U'+2(3+\zeta) U=6(2+\zeta)\, ,\label{syUD4}\\
&&W''+(1-3\zeta) W'-2(\zeta'+\zeta-\zeta^2)W= U\, ,\label{syWD4}
\ees
where 
\be\label{defYD4}
Y\equiv \frac{1}{2}W'(6-U'-2U) +W (3-6\zeta+\zeta U'+2\zeta U)+\frac{1}{4}U^2\, .
\ee
In this form the background equations explicitly show the effective dark energy density contribution $\rde=\rho_0\gamma Y$ and make $h^2$ appear only in \eq{syhD4}.
 For a constant value $\zeta_{0}$ of $\zeta$ the homogeneous solutions for $U$ and $W$ are given by
 \bees
U_{\rm hom}(x)=u_0 e^{-2x}+u_1 e^{-(3+\zeta_0)x}\, , \label{D4U}\\
W_{\rm hom}(x)=w_0 e^{2\zeta_0 x}+w_1 e^{(-1+\zeta_0)x}\, . \label{D4W}
\ees
which are constant or exponentially decreasing in MD, RD and de~Sitter epochs.
Given \eqs{syUD4}{syWD4}, during a de~Sitter inflationary stage ($\zeta_0=0$) from $x=x_{\rm in}$ to $x=x_{\rm end}$, $U$ reaches the inhomogeneous constant solution $U(x_{\rm end})\simeq 2$ while, starting from a $O(1)$ value of $W(x_{\rm in})$, at the end of inflation $W(x_{\rm end})\simeq 2\Delta N$. As pointed out when discussing the RT model, during inflation the Hubble parameter $h(x)$ becomes constant and very large and, since $V(x)=h^{-2}W(x)$, $V(x_{\rm end})$ is extremely small.

\subsection{Cosmological perturbations}

The perturbed Einstein equations are \cite{Belgacem:2017cqo}
\bees
&&\hspace*{-6mm}\( 1 - 3 \gamma \bar{V} \) \( \hat{k}^2 \Phi + 3  \Phi' -3 \Psi  \) + \frac{3 \gamma}{2} \bigg\{ \left(\bar{V}'-\frac{\bar{U}}{2h^2}\right)\delta Z+\frac{1}{2}\bar{V}' \delta Z'-
\left[3+\hat{k}^2\left(1+\frac{\bar{U}}{2}\right)\right]\delta V\nn\\
&&\hspace*{8mm}+6 \bar{V}' \Psi-3\bar{V}' \Phi'
 +\frac{\bar{U}^2}{h^2}\Phi+\left(\frac{\bar{U}'}{2}+\bar{U}-3+\frac{5}{6} \hat{k}^2 h^2 \bar{V}'\right)\delta V'-\left(\bar{U}'+2\bar{U}\right) \bar{V}' \left(\Phi+\Psi\right)
\nn\\
&& \hspace*{8mm} +\frac{2}{3}\hat{k}^2 h^2 \bar{V}'^2  \left(\Phi-\Psi\right)\bigg\}=  \frac{3}{2h^2\rho_0}\delta \rho 
\label{PoissonD4}\\
&&\hspace*{-6mm}\( 1 - 3 \gamma \bar{V} \)  \hat{k}^2 (  \Phi' - \Psi) + \frac{\gamma \hat{k}^2}{2}
\bigg[\left(\bar{U}-3\right) \left(\delta V'-\bar{V}' \Psi\right) +\bar{U} \bar{V}' \Phi\nn\\
&&\hspace*{8mm}+\frac{3}{2}\left(2-\bar{U}'-2\bar{U}-\hat{k}^2 h^2 \bar{V}'\right)\delta V
-\frac{1}{2}\bar{V}'\delta Z \bigg] =
 - \frac{3}{2 h^2\rho_0} \bar{\rho} (1+w) \hat{\theta}\label{MEE2xD4}, \\
 &&\hspace*{-6mm}
\hat{k}^2\left[( 1 - 3 \gamma \bar{V})(\Psi+\Phi)+ \gamma \bar{V}'^2 h^2 (\Psi-\Phi) 
 - 3 \gamma (1- \bar{U})\delta V +\gamma h^2 \bar{V}' \delta V'\right]\nn\\
 &&\hspace*{8mm} = \frac{9}{2 \rho_0 h^2}  e^{2x} \bar{\rho} ( 1 + w) \sigma\, ,\label{MEE4xD4}\\
&&\hspace*{-6mm}( 1 - 3 \gamma \bar{V} )\bigg[   \Phi'' + (3+\zeta) \Phi' -  \Psi' - (3+2 \zeta) \Psi + \frac{\hat{k}^2}{3}(\Phi + \Psi) \bigg] = - \frac{3}{2h^2\rho_0}\delta p
 \nn \\ 
&&\hspace*{8mm} +\frac{\gamma}{2} \bigg\{
\left[3\Phi' - \left(6+\bar{U}'+2\bar{U}+\frac{2}{3}\hat{k}^2 h^2 \bar{V}'\right)\Psi +\left(\frac{2}{3}\hat{k}^2 h^2 \bar{V}'-\bar{U}'-2\bar{U}\right) \Phi\right]  \bar{V}' \nn\\
&&\hspace*{14mm}+\left(\bar{U}-6\right)\frac{\bar{U}}{h^2}\Phi 
 + \left(\bar{V}'-\frac{\bar{U}}{2 h^2}+\frac{3}{h^2}\right)\delta Z 
 +\frac{1}{2}\bar{V}' \delta Z'\nn\\
 &&\hspace*{14mm}+ \left[3(3+2\zeta)+\left(2-\frac{\bar{U}}{2}\right)\hat{k}^2\right] \delta V
+ \left(\bar{U}'+2\bar{U}+6+\frac{5}{3}\hat{k}^2h^2\bar{V}'\right) \frac{\delta V'}{2}  \bigg\}\, ,
\label{MEE3xD4}
\ees
while the perturbation equations for the auxiliary fields are
\bees
&&\hspace*{-6mm}\delta V''+\left(1+\zeta\right)\delta V'+\bar{V}'\left(\Phi'-\Psi'\right)=
h^{-2} \left[\delta Z+ 2 \bar{U} \left(\Psi-\Phi\right)\right] \, , \label{eqVD4}\\
&&\hspace*{-6mm}\delta Z '' + \(5+\zeta\) \delta Z'+\left(6+2\zeta+2 \hat{k}^2 \right)\delta Z= -h^2 \hat{k}^4 \d V + 6 \left[\Phi''+(3+\zeta) \Phi'\right]  \label{eqZD4}\\
&&\hspace*{-4mm} + \bigg(\bar{U}'+2\bar{U}-6 - \frac{\hat{k}^2} {3} h^2 \bar{V}' \bigg) \big( \Psi' - \Phi' \big) + 2 \hat{k}^2 \big(1 - \frac{2}{3} \bar{U} \big) \Psi +4\bigg[3 \big(\zeta +2\big)+\hat{k}^2 \bigg(1 + \frac{\bar{U}}{3}\bigg) \bigg]\Phi \, .\nn
\ees
We now proceed as before, specializing these equations to a de~Sitter inflationary epoch. We consider first  the super-horizon limit, using
\eqsss{PoissonD4}{MEE4xD4}{eqVD4}{eqZD4} as independent equations,
and setting the background variables to their asymptotic values $\bar{U}=2$ and $\bar{V}=0$. Again,
we can  neglect terms involving $\bar{U}/h^2$.
Writing $\d V=v$ and $\d Z=z$ (just as in the RR case, it is not necessary to rescale the variables)
  the independent equations become
\bees
\Phi'-\Psi&=&\frac{1}{2}\gamma (v'+3v)\, ,\label{eqD4rescaled1}\\
\Phi + \Psi&=&-3\gamma v\, ,\\
v''+v'&=&0\, ,\label{eqD4rescaled2}\\
z''+5z'+6z&=&6\Phi''+20\Phi'+24\Phi-2\Psi'\, .\label{eqD4rescaled3}
\ees
Similarly to the previous discussion for the RR model, the homogeneous equations for $v$ and $z$ corresponding to \eqs{eqD4rescaled2}{eqD4rescaled3} have only constant or exponentially decreasing solutions and, as a result,  the potentials $\Phi$ and $\Psi$ are not affected by instabilities. Similar results can be obtained for $\hat{k}$ of order one. This is more easily  understood from \eqs{MEE4xD4}{MEE3xD4}, by observing that in the limit of large $h$ (and when $\hat{k}$ is not parametrically large) the terms on the right-hand sides are negligible, so the auxiliary fields
decouple from $\Psi$ and $\Phi$. Furthermore, it is easy to see that even with $\hat{k}$ generic the solutions of the homogeneous equations associated with \eqs{eqVD4}{eqZD4} are stable.
Finally,
in the opposite limit $\hat{k}\gg 1$, all time derivatives in \eqst{PoissonD4}{eqZD4} are suppressed with respect to terms involving $\hat{k}^2$ or $\hat{k}^4$, and the solutions for $\Phi, \Psi, v$ and $z$ are static and suppressed by powers of $1/\hat{k}^2$ or $1/h^2$, so again there is no instability.
Therefore the $\Delta_4$ model provides a stable  cosmological evolution also during inflation, for both super-horizon and sub-horizon modes.

\section{Conclusions}

In a theory with massless  particles (such as the graviton in GR), quantum corrections unavoidably induce non-local terms in the quantum effective action. In the ultraviolet regime the computation of these non-local terms is well understood theoretically, and is by now standard textbook material (see e.g. \cite{Birrell:1982ix,Barvinsky:1985an,Barvinsky:1987uw,Buchbinder:1992rb,Mukhanov:2007zz,Shapiro:2008sf}). The non-local terms computed in the ultraviolet regime of GR, however, are only relevant at Planckian energy scales. For cosmological applications, we are rather interested in the infrared regime. In this case the situation is much less understood theoretically. In general, there are several hints that large infrared effects can appear.  For instance, because of strong infrared fluctuations, a  massless minimally-coupled scalar field in de~Sitter space   develops  dynamically a mass   \cite{Starobinsky:1994bd,Riotto:2008mv,Burgess:2009bs,Rajaraman:2010xd,Beneke:2012kn,Gautier:2013aoa}. In the case of pure gravity the  existence of strong IR effects in gauge-invariant quantities in de~Sitter space has been the subject of many investigations, with  conflicting results (see e.g. \cite{Miao:2011ng,Rajaraman:2016nvv} for  discussions and review of the literature), and it is fair to say that the infrared limit of quantum gravity  is currently not yet well understood.

Given this state of affairs, it makes sense to take a phenomenological attitude, and explore whether some nonlocal terms relevant in the IR, and associated with a mass scale $m$ (which is understood to be generated dynamically) could have interesting cosmological consequences. The present paper is part of an effort of our group in this direction (see \cite{Maggiore:2016gpx,Belgacem:2017cqo} for comprehensive reviews). The requirement of obtaining a viable cosmological evolution, both at the background level and at the level of cosmological perturbations, severely restricts the class of nonlocal terms that can be considered. For instance, terms involving $\iBox\RMN$ in the equations of motion, as well as terms in the action of the form $\RMN\Box^{-2}\Rmn$, or similar terms involving the Weyl tensor, do not work~\cite{Maggiore:2013mea,Foffa:2013vma,Cusin:2015rex}. Basically, this restricts the attention to terms involving nonlocal operators acting on the Ricci scalar, as in the RT, RR and $\Delta_4$ models. This is quite interesting conceptually, since the nonlocal terms in these  models,  in the linearized limit, correspond to introducing a nonlocal but diffeomorphism-invariant   mass term for the conformal mode \cite{Maggiore:2015rma,Maggiore:2016fbn,Maggiore:2016gpx}.

 In this paper we have discussed whether a further constraint on the viable models emerges  by requiring stability of the background evolution and of the cosmological perturbations even during a phase of primordial inflation, while in previous works we only required stability in RD and MD. We find that  the evolution of the RR and $\Delta_4$ models is stable. For the RT model there is an apparent instability, which has however completely negligible effects because, during primordial inflation, the variables in terms of which there is an apparent instability have initial conditions suppressed by a huge factor,
given by the ratio of the Hubble parameter during inflation and that today, squared, and therefore, even if they  grow, they still remain at extremely small values. Thus,
also the RT model passes this test. This provides useful information in view of attempts at deriving these nonlocal terms in the quantum effective action from a fundamental action.   

\vspace{5mm}

\noindent
{\bf Acknowledgments.} 
The work of E. Belgacem, G. Cusin, S. Foffa and M. Maggiore is supported by the Fonds National Suisse and  by the SwissMap NCCR. We thank Filippo Vernizzi for useful discussions.

\bibliographystyle{utphys}
\bibliography{myrefs_massive}

\end{document}

%% file: mydefs.tex
\newcommand{\DP}{\Delta_4}

\newcommand{\nn}{\nonumber}

\newcommand{\iBox}{\Box^{-1}}

\renewcommand\({\left(}
\renewcommand\){\right)}
\renewcommand\[{\left[}
\renewcommand\]{\right]}

\newcommand\n{{\mbox {\boldmath $\nabla$}}}
\newcommand{\ra}{\rightarrow}

\def\lsim{\raise 0.4ex\hbox{$<$}\kern -0.8em\lower 0.62
ex\hbox{$\sim$}}

\def\gsim{\raise 0.4ex\hbox{$>$}\kern -0.7em\lower 0.62
ex\hbox{$\sim$}}

\def\lbar{{\hbox{$\lambda$}\kern -0.7em\raise 0.6ex
\hbox{$-$}}}

\newcommand\eq[1]{eq.~(\ref{#1})}
\newcommand\eqs[2]{eqs.~(\ref{#1}) and (\ref{#2})}

\newcommand\eqsss[4]{eqs.~(\ref{#1}), (\ref{#2}), (\ref{#3})
and (\ref{#4})}

\newcommand\eqst[2]{eqs.~(\ref{#1})--(\ref{#2})}

\newcommand\pa{\partial}
\newcommand\p{\partial}

\newcommand\ee{\end{equation}}
\newcommand\be{\begin{equation}}
\def\bea{\begin{array}}
\def\eea{\end{array}}\def\ea{\end{array}}
\newcommand\ees{\end{eqnarray}}
\newcommand\bees{\begin{eqnarray}}
\def\nn{\nonumber}

\def\a{\alpha}

\def\g{\gamma}

\def\d{\delta}

\def\dslash{\hspace{-1mm}\not{\hbox{\kern-2pt $\partial$}}}
\def\Dslash{\not{\hbox{\kern-2pt $D$}}}
\def\pslash{\not{\hbox{\kern-2.1pt $p$}}}
\def\kslash{\not{\hbox{\kern-2.3pt $k$}}}
\def\qslash{\not{\hbox{\kern-2.3pt $q$}}}

\newcommand{\vx}{{\bf x}}

\def\p1{{\bf p}_1}
\def\p2{{\bf p}_2}
\def\k1{{\bf k}_1}
\def\k2{{\bf k}_2}

\newcommand{\gmn}{g_{\mu\nu}}

\newcommand{\gMN}{g^{\mu\nu}}

\newcommand{\pan}{\pa_{\nu}}

\newcommand{\Rmn}{R_{\mu\nu}}
\newcommand{\Gmn}{G_{\mu\nu}}
\newcommand{\RMN}{R^{\mu\nu}}

\newcommand{\Tmn}{T_{\mu\nu}}

\newcommand{\dddM}{\kern 0.2em \raise 1.9ex\hbox{$...$}\kern -1.0em \hbox{$M$}}
\newcommand{\dddQ}{\kern 0.2em \raise 1.9ex\hbox{$...$}\kern -1.0em \hbox{$Q$}}
\newcommand{\dddI}{\kern 0.2em \raise 1.9ex\hbox{$...$}\kern -1.0em\hbox{$I$}}
\newcommand{\dddJ}{\kern 0.2em \raise 1.9ex\hbox{$...$}\kern-1.0em
\hbox{$J$}}
\newcommand{\dddcalJ}{\kern 0.2em \raise 1.9ex\hbox{$...$}\kern-1.0em
\hbox{${\cal J}$}}

\newcommand{\dddO}{\kern 0.2em \raise 1.9ex\hbox{$...$}\kern -1.0em
\hbox{${\cal O}$}}
\def\dddz{\raise 1.5ex\hbox{$...$}\kern -0.8em \hbox{$z$}}
\def\dddd{\raise 1.8ex\hbox{$...$}\kern -0.8em \hbox{$d$}}
\def\dddbd{\raise 1.8ex\hbox{$...$}\kern -0.8em \hbox{${\bf d}$}}
\def\ddbd{\raise 1.8ex\hbox{$..$}\kern -0.8em \hbox{${\bf d}$}}
\def\dddx{\raise 1.6ex\hbox{$...$}\kern -0.8em \hbox{$x$}}

\newcommand{\mplr}{m_{\rm Pl}}

\newcommand{\rde}{\rho_{\rm DE}}
\newcommand{\wde}{w_{\rm DE}}
